\documentclass[universe,article,accept,pdftex,moreauthors]{Definitions/mdpi}

\firstpage{1}
\pubvolume{1}
\issuenum{1}
\articlenumber{0}
\pubyear{2024}
\copyrightyear{2024}
\datereceived{5 January 2024}
\daterevised{19 February 2024} 
\dateaccepted{23 February 2024}
\datepublished{ }
\hreflink{https://doi.org/} 

\usepackage{aas_macros,float}
\usepackage{graphicx,epstopdf,amsmath,amssymb,amsthm,euscript,endnotes,color, tipa, dsfont}

\Title{Search for Wormhole Candidates: Accreting Wormholes with Monopole Magnetic Fields}

\TitleCitation{Search for Wormhole Candidates: Accreting Wormholes with Monopole Magnetic Fields}

\Author{{Mikhail Piotrovich} 
 {*}\href{https://orcid.org/0000-0002-0838-7516}{\orcidicon}, Serguei Krasnikov \href{https://orcid.org/0000-0002-9320-4189}{\orcidicon}, Stanislava Buliga \href{https://orcid.org/0000-0003-0182-5058}{\orcidicon} and Tinatin Natsvlishvili \href{https://orcid.org/0000-0002-4627-511X}{\orcidicon}}

\AuthorNames{Mikhail Piotrovich, Serguei Krasnikov, Stanislava Buliga and Tinatin Natsvlishvili}

\AuthorCitation{Piotrovich, M.; Krasnikov, S.; Buliga, S.; Natsvlishvili, T.}

\address[1]{%
Central Astronomical Observatory at Pulkovo RAS, 196140 Saint-Petersburg, Russia; {{krasnikov.xxi@gmail.com}
(S.K.); aynim@yandex.ru (S.B.); tinatingao@mail.ru (T.N.)}}

\corres{\hangafter=1 \hangindent=1.05em \hspace{-0.82em}Correspondence: mpiotrovich@mail.ru}

\abstract{The existence of even the simplest magnetized wormholes may lead to observable consequences. In the case where both the wormhole and the magnetic field around its mouths are static and spherically symmetric, and gas in the region near the wormhole falls radially into it, the former's spectrum contains bright cyclotron or synchrotron lines due to the interaction of charged plasma particles with the magnetic field. At the same time, due to spherical symmetry, the radiation is non-polarized. The emission of this just-described exotic type (non-thermal, but non-polarized) may be a wormhole signature. Also, in this scenario, the formation of an accretion disk is still quite possible at some distance from the wormhole, but a monopole magnetic field could complicate this process and lead to the emergence of asymmetrical and one-sided relativistic jets.}


\newcounter{aaa}
{\end{trivlist}}

\newcommand{\rmd}{\mathrm{d}}

\newcommand*{\bi}[1]{\boldsymbol{#1}}

\newcommand{\x}{\mathcal{ X}}%

\begin{document}

\section{Introduction}

\subsection{Wormholes}

According to the ``boring physics conjecture'' \citep{visser95}, we live in $\mathds{R}^4$ or, at best, in $\mathds{R}\times\mathds{S}^3$. On the other hand, \citet{kardashev07} proposed the hypothesis that some galactic nuclei are, in fact, wormhole mouths (see also \citet{bambi13,li14,zhou16}). Evidently, the time is not ripe to discuss the topology of the Universe purely theoretically.

The study of wormholes is of serious interest since their properties and the very possibility of their existence can have a strong impact on our ideas about the cosmology of the Universe.

Wormholes (also known as ``Einstein-Rosen bridges'') were first proposed by Einstein and Rosen \citep{einstein35} within the framework of general relativity. The Einstein--Rosen bridge solution describes an empty, spherically symmetric wormhole geometry that connects two asymptotically flat regions of spacetime. These hypothetical objects are essentially shortcuts through spacetime, connecting distant regions of the Universe or even different universes. This idea has generated great interest among scientists, inspiring many fascinating theories and proposals.

The solutions to the equations of general relativity allow for the existence of traversable and non-traversable (those that collapse too soon to be traversed) wormholes depending on the energy-matter content of spacetime. It should be noted that the theory of traversable wormholes, which could theoretically allow for fast interstellar travel, has a lot of constraints and challenges, including the requirement for negative energy density ``exotic'' matter that should stabilize the wormhole throat in order to prevent its collapse. Non-traversable wormholes also have important implications for theoretical physics and cosmology, allowing us to test the limits of general relativity and study the nature of spacetime under extreme~conditions.

One of the serious problems in the study of wormholes is the preservation of causality. Traversable wormholes could allow time travel, which could lead to apparent paradoxes \citep{einstein49}. One such paradox is the classic ``grandfather paradox''. Imagine a scenario where a man, equipped with the ability to travel through time, makes a fateful decision to eliminate his own grandfather during infancy. With resolve, he journeys into the past, sneaks up on the helpless infant, and takes a shot. What unfolds next? The situation seems to spiral into a paradox. On one hand, the baby is indeed killed. Yet, on the other hand, the act cannot come to pass, for if the grandfather perishes, the father of the time-traveler—and consequently the time-traveler himself—would never come into existence, leaving no one to carry out the fatal deed. Theoretical approaches have been proposed that allow spacetime to avoid paradoxes \citep{krasnikov02}. ``The Chronology Protection Conjecture'', proposed by Stephen Hawking \citep{hawking92}, suggests that in order to preserve causality, the laws of physics prevent the formation of closed time-like curves. The stability and consistency of wormholes within the framework of general relativity continue to be actively studied.

There are many authors who have tried to determine the unique observational features of wormholes \citep{harko08,harko09,tsukamoto12,yoo13,bambi13,bambi13b,dokuchaev14,li14,zhou16,dai19,paul20,tripathi20,bambi21}, in particular, the features that distinguish wormholes from black holes. However, almost all of them have concluded that these features, even when they exist, are indeed difficult (and in some cases, simply impossible) to observe with modern astronomical equipment. In our work, we try to use objects and physical mechanisms that are well known in astrophysics, allowing us to obtain strong observational~manifestations.

\subsection{Accretion}

Accretion is a very important phenomenon, occurring in various astrophysical objects such as normal stars and compact relativistic objects (including active galactic nuclei). The understanding of accretion processes is important for explaining the observed properties of these objects, such as their luminosity, spectral characteristics, and dynamics. In many astrophysical scenarios, magnetic fields play a significant role in shaping the flows of accreting matter. The interaction between magnetic fields and accreting matter adds additional complexity to the accretion process, leading to various phenomena such as the formation of magnetized accretion disks \citep{bardeen71} and collimated jets \citep{blandford77,blandford82} and the emergence of magnetic turbulence.

It is generally accepted that active galactic nuclei (AGNs) and quasars (QSOs) often possess a magnetized accretion disk \citep{blaes04,moran08}. There are many models of accretion disk structures. The most commonly used and well-known model is the standard Shakura--Sunyaev model \citep{shakura73}. In this model, the disk is held vertically by thermal pressure and turbulent viscosity is used to explain the transfer of angular momentum required by the accretion flow. Shakura--Sunyaev's model is simple and convenient, but it has a number of problems. For example, this model does not predict X-ray emission in AGN spectra, and recent observations have shown that the size of the accretion disk is several times larger than that predicted by the Shakura--Sunyaev model (see, for example, \citet{fausnaugh16}).

In addition, there are other models. We will list just a few of them. For example, \citet{balbus98} showed that accretion disks have a robust mechanism for generating magnetohydrodynamic (MHD) turbulence due to magnetorotational instability. \citet{miller00} studied disks with initial Gaussian density profiles supported by thermal pressure. In the case of an initial axial magnetic field, \citet{miller00} observed that the saturated magnetic pressure dominates thermal pressure not only in the corona but also everywhere in the disk. Previously, \citet{eardley75} and \citet{field93} considered analytic models of thin accretion disks with angular momentum transfer due to magnetic stresses. Both these works included magnetic loops with sizes of the order of the disk thickness. \mbox{\citet{shalyabkov00,campbell00}, and \citet{ogilvie01}} studied models of magnetized accretion disks with an externally imposed large-scale vertical magnetic field and anomalous magnetic field diffusion due to enhanced turbulent diffusion. In general, it can be said that studying accretion processes in the presence of magnetic fields is very important for advancing our understanding of astrophysical phenomena.

\subsection{Magnetic Fields}

Magnetic fields in the Universe are generated by various processes, such as dynamo action in stellar interiors, the amplification of primordial magnetic fields in the early Universe, and the compression and stretching of magnetic fields in accretion disks. Magnetic fields are typically classified based on their origin, strength, and topology. There are, for example, dipole, quadrupole, and toroidal fields. Monopole magnetic fields, which have isolated magnetic poles, are a hypothetical possibility that could have important consequences for astrophysics and fundamental physics if detected. The importance of the monopole field is that it directly connects the properties of experimentally observable fields with the topology of spacetime.

\subsection{Wormholes with Monopole Magnetic Fields}

Due to their unique properties, wormholes provide an interesting opportunity to use them as an object that can theoretically generate a monopole magnetic field. And, there are many works linking wormholes with a monopole field \citep{misner57,bronnikov73,ellis73,agnese96,prat-camps15,romero15,romero17,romero19,canate23}.

In the pioneering work of Misner and Wheeler \citep{misner57}, the concept of traversable wormholes was introduced, and it was also pointed out that they might have magnetic and scalar fields (``charge without charge''). After that, Bronnikov \citep{bronnikov73} and Ellis \citep{ellis73} considered a specific solution of Einstein's equation where wormholes are supported by electromagnetic and scalar fields.

\citet{agnese96} demonstrated that a Kaluza--Klein theory in five dimensions, derived through a conformal gauge approach called the ``total space'', can depict spacetimes that accommodate both magnetic monopoles and wormhole structures. This means that within this theoretical framework, it is possible to describe regions of space where magnetic charges exist alongside passages connecting different points in spacetime.

\citet{prat-camps15} used an interesting approach. While the idea of creating a wormhole in a laboratory may seem daunting, they used an unusual theoretical approach to construct a wormhole for electromagnetic waves using metamaterials. This would enable the transmission of electromagnetic waves through an invisible tunnel between two points in space. The researchers successfully built and demonstrated a magnetostatic wormhole. By utilizing magnetic metamaterials and metasurfaces, their wormhole was capable of transferring magnetic fields from one spatial point to another without detection. Experimental results illustrated that the magnetic field generated at one end of the wormhole appeared at the other end as an isolated monopole magnetic field. This created the illusion of a magnetic field traversing through a tunnel outside of conventional three-dimensional space.

{\citet{romero15} applied the Weitzeb{\"o}ck-Induced Matter Theory (WIMT) to two specific metrics: the Gullstr{\"a}nd–Painlev{\'e} and Reissner–Nordstr{\"o}m metrics.} 
This method is a recent development that expands upon the Induced Matter Theory by utilizing Weitzeb{\"o}ck's geometry on a curved 5D manifold. The key insight exploited here is that the Riemann–Weitzenb{\"o}k curvature tensor is consistently null. Through this approach, the study revealed the existence of currents, the interpretation of which suggests the potential presence of stable gravito-magnetic monopoles.

\citet{romero17} utilized the Weitzenb{\"o}ck-Induced Matter Theory to analyze Schwarzschild wormholes within an extended 5D manifold, which included non-vacuum conditions. They investigated various ways of describing the wormholes, known as foliations, and examined the geodesic equations governing the motion of observers situated within a traversable wormhole. Additionally, the study explored how these observers could detect gravito-magnetic monopoles, which are essentially the gravitational analogs of magnetic monopoles, and contrasted them with gravito-electric sources typically observed in the outer region of Schwarzschild black holes. The researchers also calculated the densities of these monopoles and discussed their quantization according to the principles outlined by Dirac. Their analysis revealed a duality within the extended Einstein--Maxwell equations, linking electric and magnetic charges across spatial regions that are causally~disconnected.

\citet{romero19} focused on examining a traversable wormhole generated through a transformation applied to the 4D Dymnikova metric, which characterizes analytical black holes. The study employed a coordinate transformation method inspired by the Einstein--Rosen bridge to analyze a particular set of geodesics. These geodesics involved test particles carrying electric charges, which, due to the transformation, induced an effective magnetic monopole that could be observed by external observers situated outside the wormhole. Given that traditional Riemannian geometry does not account for the existence of magnetic monopoles, the study introduced torsional geometry as a potential explanation for the geometric induction of magnetic monopoles. The researchers derived an equation linking torsion and magnetic fields, along with a mathematical expression similar to Dirac's equation that describes magnetic and electric charges. This formulation suggests that torsion could give rise to a fundamental length scale, enabling the generation of a magnetic field and introducing a discretization of spacetime.

\citet{canate23} described the discovery of magnetically charged ultrastatic and spherically symmetric spacetime solutions within the framework of both linear and nonlinear electrodynamics, coupled with Einstein-scalar-Gauss--Bonnet (EsGB-$\mathcal{L}(\mathcal{F})$) gravity. These solutions are characterized by an electromagnetic Lagrangian density $\mathcal{L}(\mathcal{F})$, which solely depends on the electromagnetic invariant $\mathcal{F} = F_{\alpha\beta}F^{\alpha\beta} /4$. The paper highlighted a particular class of these solutions, where the electromagnetic invariant $\mathcal{F}$ attains a strict global maximum value $\mathcal{F}_{0}$ across the entire solution domain, and the Lagrangian density $\mathcal{L}_{0}=\mathcal{L}(\mathcal{F}_{0})>0$. It was shown that such solutions can be interpreted as ultrastatic wormhole spacetime geometries, with the radius of the wormhole throat determined by the scalar charge and the quantity $\mathcal{L}_{0}$. Examples provided included Maxwell's theory of electrodynamics (linear electrodynamics) with $\mathcal{L}_\mathrm{LED} = \mathcal{F}$, which yields the magnetic dual of the purely electric Ellis--Bronnikov EsGB Maxwell wormhole. Additionally, nonlinear electrodynamics (NLED) models, such as Born--Infeld $\mathcal{L}_{\mathrm{BI}} = -4\beta^{2} + 4\beta^{2} \sqrt{ 1 + \mathcal{F} /(2\beta^{2})~}$ and Euler--Heisenberg in the weak-field limit $\mathcal{L}_\mathrm{EH} = \mathcal{L}_\mathrm{LED} + \gamma \mathcal{F}^2 /2$, were discussed. Using these NLED models, two new magnetically charged ultrastatic traversable wormholes (EsGB Born--Infeld and EsGB Euler--Heisenberg wormholes) were presented as exact solutions within EsGB-$\mathcal{L}(\mathcal{F})$ gravity. These solutions do not require exotic matter, and it was demonstrated that they share the characteristic that, in the weak electromagnetic field region, the magnetically charged Ellis--Bronnikov EsGB Maxwell wormhole is recovered.

\subsection{Polarization}

Let us talk in more detail about the mechanisms for generating polarized radiation. Polarization, being sensitive to the anisotropy of the matter distribution, plays a crucial role in the study of optically unresolved central regions of AGNs, such as the accretion disk. An accretion disk is a typical example of a radiating region with a non-spherically symmetric electron density distribution. As a result of scattering by plasma electrons, the disk radiation becomes polarized. {Polarimetric observations indicate that AGNs and QSOs have polarized emissions across a variety of wavelength ranges, from ultraviolet to radio waves, in continuous-wave and linear emissions \citep{martin83,impey95,wilkes95,barth99,smith02,modjaz05}.} 
These works discuss that the observed polarization has different mechanisms of origin: light scattering in accretion disks, which occurs on both free and bound electrons, and cyclotron and synchrotron radiation of charged particles. These mechanisms can work within different structures, such as plane and warped accretion disks, as well as toroidal clumpy rings surrounding the accretion disks and jets. Often, different models are proposed to explain the same source. There are many works devoted to the study of various aspects of the structure and radiation of AGNs and quasars. Of particular interest is the mechanism of the generation of relativistic jets. According to modern concepts, a large-scale magnetic field plays a key role in the launch of relativistic jets~\citep{blandford77,blandford82,lovelace87}, and its toroidal component effectively collimates the jets \citep{benford78,chan80}. The magnetic field manifests itself in linearly polarized synchrotron radiation and the Faraday rotation effect. In BL Lac objects (BL Lacs), the electric vector position angle often coincides with the local direction of the jet \citep{gabuzda00,lister05,osullivan09}, while quasars exhibit a distribution without a preferred direction \citep{lister05}.

\subsection{Our Specific Approach}

Previously, we considered the possible accretion of matter into a wormhole \citep{piotrovich20a,piotrovich20b}. Specifically, we studied the case where accretion into a traversable wormhole occurs from both sides, each of which is located at the center of the active galactic nucleus. As a result, high-energy accretion flows collide inside the wormhole, which can lead to plasma heating to extremely high temperatures of up to $10^{14}$K. Plasma with such parameters would exhibit a very specific spectrum, distinct from that of ordinary active galactic nuclei.

Now, we consider accretion in the presence of a magnetic field, assuming that the wormhole has a monopole magnetic field, which greatly distinguishes it from the much better-studied Kerr black hole.

In this paper, we argue that the existence of even the simplest magnetized wormholes may lead to observable consequences not yet discussed (to the best of our knowledge) in the literature. Indeed, consider the case where both the wormhole and the magnetic field around its mouth are static and spherically symmetric. Suppose that gas in the region near the wormhole \mbox{falls radially} into it Then, the former's spectrum contains bright cyclotron (or, in relativistic cases, synchrotron) lines due to the interaction of the charged plasma particles with the magnetic field. At the same time, due to spherical symmetry, the radiation is non-polarized. This is a rather unusual combination since in known astrophysical objects, cyclotron and synchrotron radiation is caused, as a rule, by a dipole-like magnetic field, which is not spherically symmetric. In addition, synchrotron radiation can be generated, for example, by relativistic jets, the geometry of which also leads to strong~polarization.

Thus, we can speculate that the emission of the just-described exotic type (cyclotron or synchrotron, but non-polarized) may be a wormhole signature.

\textls[-25]{To convey the essence of this phenomenon we also show the possible trajectories of a charged particle near a wormhole using numerical simulations, under certain simplifying~assumptions}.

Very little is known about wormholes, including their birth and evolution. We try to compensate for this by applying the most general arguments to the simplest astrophysical monopole (the meaning of the word ``simplest'' is assumed to be intuitively clear). In particular, they must be static and spherically symmetric.

Although theoretically, the monopole object can be, for example, a magnetically charged Reissner--Nordstr\"om black hole with its horizons, singularities, and an infinite set of asymptotically flat ends, it is by no means simple. So, we do not consider such a case.

\section{Our Model and Some Calculations}

\subsection{Toy Model and Basic Equations}

The simplest compact radially magnetized object can be a wormhole based on the Reissner--Nordstr\"om spacetime (actually, as long as the magnetic field outside the object is a monopole, the structure of the former is irrelevant).

Pick three positive parameters: $Q;\ m$, where $m>Q;$  and $ r_0$, where the former two describe the magnetic charge and ``mass'' of the wormhole, respectively, and $r_0$ obeys the {inequality} 

\[
    r_0 >r_\text{Horizon} ,\quad    r_\text{Horizon}\equiv m + \sqrt{m^2 - Q^2}
\]
\textls[-15]{and characterizes ``the size'' of the  wormhole. The auxiliary ``half-wormhole'', $W_1$, is defined~{as} 
}

\begin{equation}\label{eq:}
    W_1:\qquad \rmd s^2= -\nu(r) \,  \rmd t^2 + \nu(r)^{-1}\rmd r^2  + r^2 (\rmd\theta^2 + \cos^2\theta\,\rmd\phi),
\end{equation}

\noindent where

\begin{equation}\label{RNeq:}
    r > r_0, \qquad t \in \mathds{R}, \qquad \nu \equiv  1 - \tfrac{2m}{r} + \tfrac{Q^2}{r^2}.
\end{equation}

$W_1$ is \emph{almost} the sought-after spacetime: it is static, spherically symmetric, and when endowed with the radial magnetic field

\begin{equation}\label{eq:B}
    \bi B= \Big (Q/r^2\Big ){\bi e}_{\hat r}
\end{equation}

\noindent solves the Maxwell--Einstein equations. The only ``drawback'' of $W_1$ is that it has a single asymptotically flat end, so $W_1$ is an extendable---where one of its extensions is denoted as $U_1$---funnel rather than a wormhole. To eliminate this drawback, we define the wormhole, $W$, as a pair of equal funnels, $U_1$ and $U_2$, with their stems identified (the existence of a suitable isometry relating regions $U_1$ and $U_2$ is a non-trivial condition), as illustrated in Figure~1 in the work by \mbox{\citet{morris88a}}.

\subsection{Interaction of Accreting Matter with a Monopole Magnetic Field}

A detailed description of the behavior of a plasma flow in the presence of a monopole magnetic field is an extremely difficult task and is beyond the scope of the current rather phenomenological study. And although in this work, we primarily consider spherical accretion, one can reasonably assume that a monopole field would impede the movement of plasma along Keplerian orbits near the wormhole and, in particular, complicate the formation of an accretion disk. However, the formation of an accretion disk is still quite possible at some distance from the wormhole, where the low magnetic field and low temperature (and, therefore, low degree of ionization) of the accreting matter would not allow the monopole field to strongly influence the matter. It should be noted that accretion disks near wormholes without their own magnetic fields have already been considered in the literature \citep{harko09,lobo17,piotrovich20b}.

We can also purely phenomenologically study the formation of relativistic jets by the accretion disk at a wormhole with a monopole magnetic field. Since the disk wind would be suppressed by a monopole field, the formation of a jet is possible, most likely only through the Blandford--Znajek mechanism \citep{blandford77}, in which the surrounding interstellar matter (not from the disk) is collimated due to the interaction of the disk's poloidal magnetic field with the rotating black hole (in our case, the wormhole). Thus, if the wormhole rotates, jets can, in principle, form. Moreover, if the monopole field of a wormhole is much stronger than the dipole field of the accretion disk, the cyclotron radiation from them would be significantly stronger and at other frequencies compared to the black hole case. It is also possible that the jets themselves would be more powerful/faster, but this is not certain. If the strength of the wormhole's monopole field is comparable to the disk's dipole field, as a result of the superposition of fields at one of the poles of the dipole field, there would be a sharp decrease in the field's strength. As a result, firstly, the cyclotron radiation from this side of the jet would be significantly weaker, and, secondly, the jet itself may be less powerful, less collimated, or even not formed at all (but this, again, is not certain; the mechanism of jet formation is quite complex and is still not fully understood). It should be noted that asymmetrical and one-sided jets have actually been observed near some active galactic nuclei \citep{bridle84,parma87,cawthorne91}, and this fact is rather difficult to explain using classical theories, whereas our wormhole model offers a simple explanation.

\subsection{Numerical Simulation}

Let us consider a simpler problem, namely the movement of one charged particle near a gravitating object with a monopole magnetic field in the simplified fictional space $\mathds{R}^3$ (which is the Newtonian approximation of our metric $W$), where we use the Newton gravity approximation (here, the ``mass'' and the charge of the wormhole are described by the parameters $m$ and $Q$, and it is a point-like object at the origin). In this case, the trajectory of the particle moving far enough from a gravitating object (so that relativistic effects can be neglected as a first approximation) can be obtained relatively easily (in our model, we use non-relativistic Maxwell equations). Let us consider, for example, a proton near a point-like object with the mass of the Sun $M_\odot$. In order to define the magnetic field strength, we set the value of the field $B_{10}$ at a radius of $R = 10$km. For simplicity, we neglect the loss of proton energy due to cyclotron radiation since the trajectory is built on a time interval of only 0.01 s.

We have numerically calculated the proton trajectories for various values of parameters such as the magnetic field strength $B_{10}$, starting position $R_\text{st}$, and velocity $V_\text{st}$ of a proton using a rather simple method. The force acting on the proton consists of the Lorentz force and the gravitational force $\vec{F} = \vec{F_L} + \vec{F_g} = q_p / c [\vec{V} \times \vec{B}] - G M_\odot m_p \vec{r}/r^3$, where $q_p$ is the proton charge, $c$ is the speed of light, $\vec{V}$ is the proton speed, $G$ is the gravitational constant, $m_p$ is the proton mass, $\vec{r}$ is the radius vector, $\vec{B} = \mu \vec{r} / r^3$ is the magnetic monopole field strength, and $\mu$ is the magnetic permeability, which, in our case, is $\mu = B_{10} R^2$.

In this calculation, for convenience, we measure the distance in centimeters, the mass in grams, and the time in $10^{-8}$s. First, we set the starting position $x_0$, $y_0$, $z_0$ and starting velocity $\dot{x_0}$, $\dot{y_0}$, $\dot{y_0}$ in the Cartesian coordinate system. Then, at each iteration, the acceleration of the proton due to the Lorentz force and the Newtonian force of attraction are calculated as~follows:

\begin{equation}
\begin{aligned}
    r & = (x^2 + y^2 + z^2)^{1/2},\\
    \ddot{x} & = ((y \dot{z} - \dot{y} z) (10^8 B_{10}) - x G M_\odot) / r^3,\\
    \ddot{y} & = ((z \dot{x} - \dot{z} x) (10^8 B_{10}) - y G M_\odot) / r^3,\\
    \ddot{z} & = ((x \dot{y} - \dot{x} y) (10^8 B_{10}) - z G M_\odot) / r^3.
    \label{forces}
\end{aligned}
\end{equation}

After that, the change in speed due to acceleration and the change in coordinates due to speed are calculated as follows:

\begin{equation}
\begin{aligned}
    \dot{x} & \rightarrow \dot{x} + \ddot{x} dt,\\
    \dot{y} & \rightarrow \dot{y} + \ddot{y} dt,\\
    \dot{z} & \rightarrow \dot{z} + \ddot{z} dt,\\
    x & \rightarrow x + \dot{x} dt,\\
    y & \rightarrow y + \dot{y} dt,\\
    z & \rightarrow z + \dot{z} dt,
    \label{iteration}
\end{aligned}
\end{equation}

\noindent where in this case, the optimal value of time step $dt$ turned out to be 0.001 in our units.

Thus, $10^9$ iterations were performed for the proton flight time of 0.01 s. If the proton flew closer to the central object than 10 km, we considered that it would inevitably fall on the object and the calculation would be stopped. To avoid the accumulation of errors, we ensured that the total (kinetic plus potential) energy of the proton divided by the proton mass, $E_p / m_p = (\dot{x}^2 + \dot{y}^2 + \dot{z}^2) / 2.0 - G M_\odot / \sqrt{(x^2 + y^2 + z^2)}$, did not change by more than 0.001.

\begin{figure}[H]
  \includegraphics[bb= 85 20 705 570, clip, width=0.58\textwidth]{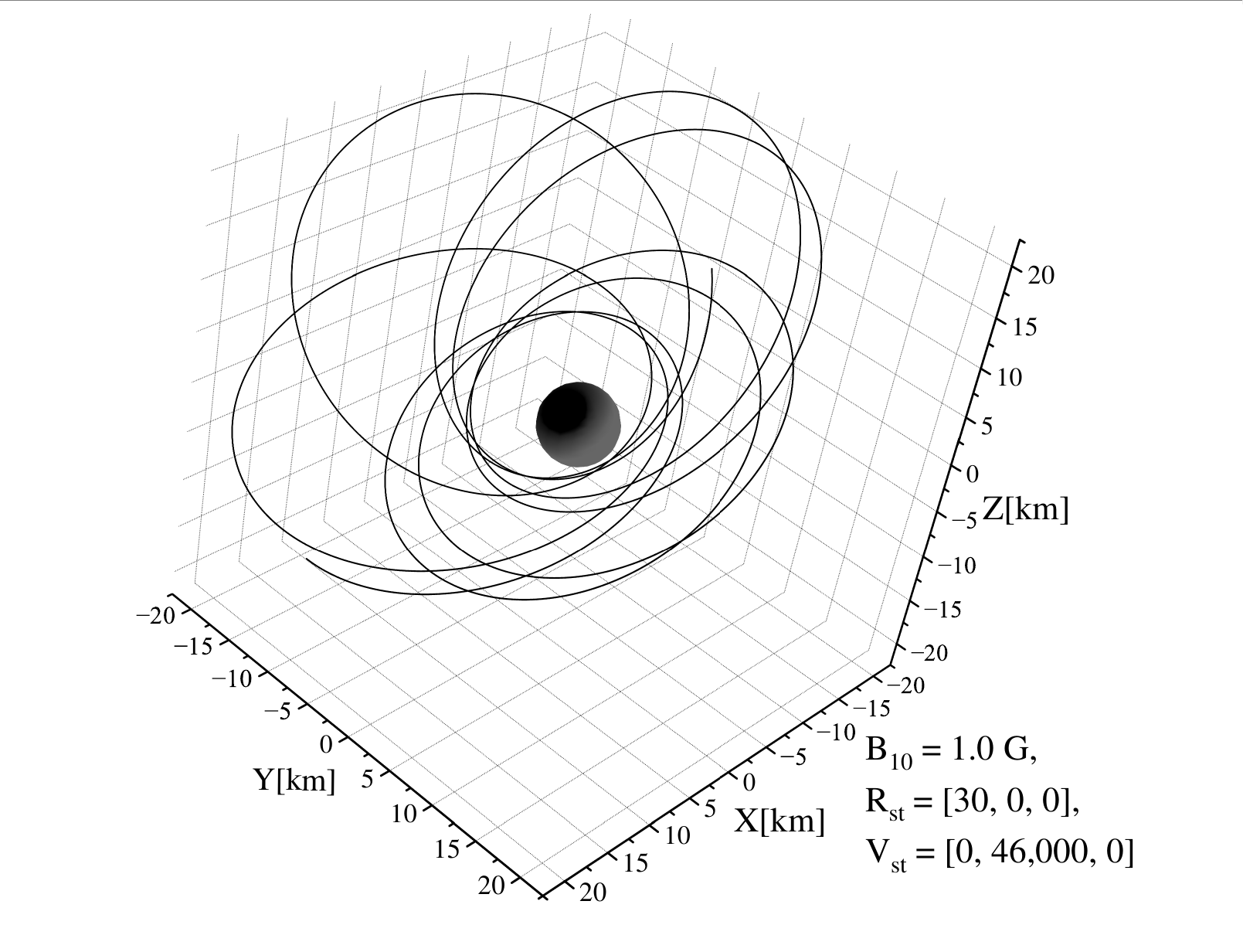}\\
  \includegraphics[bb= 85 20 705 570, clip, width=0.58\textwidth]{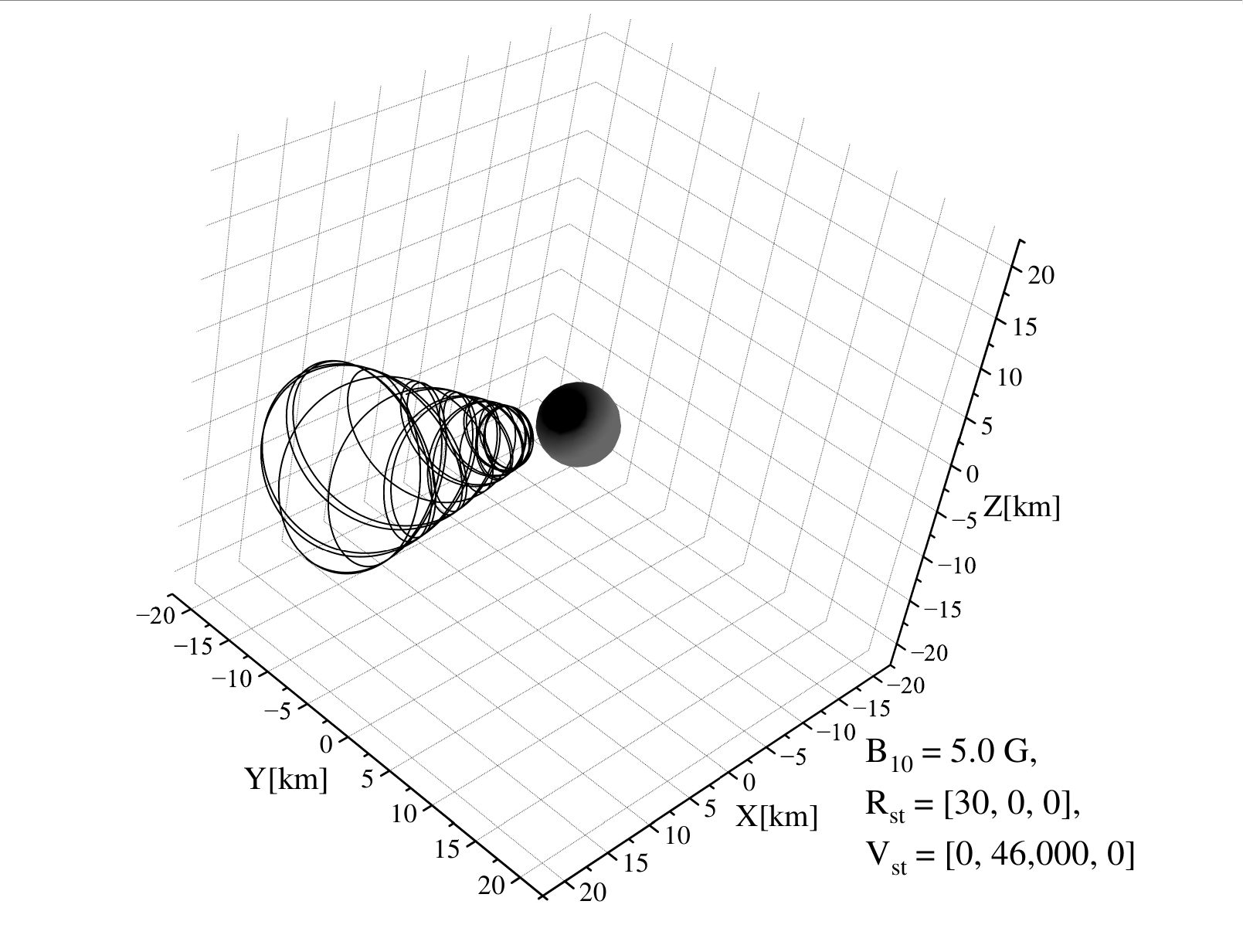}\\
  \includegraphics[bb= 85 20 705 570, clip, width=0.58\textwidth]{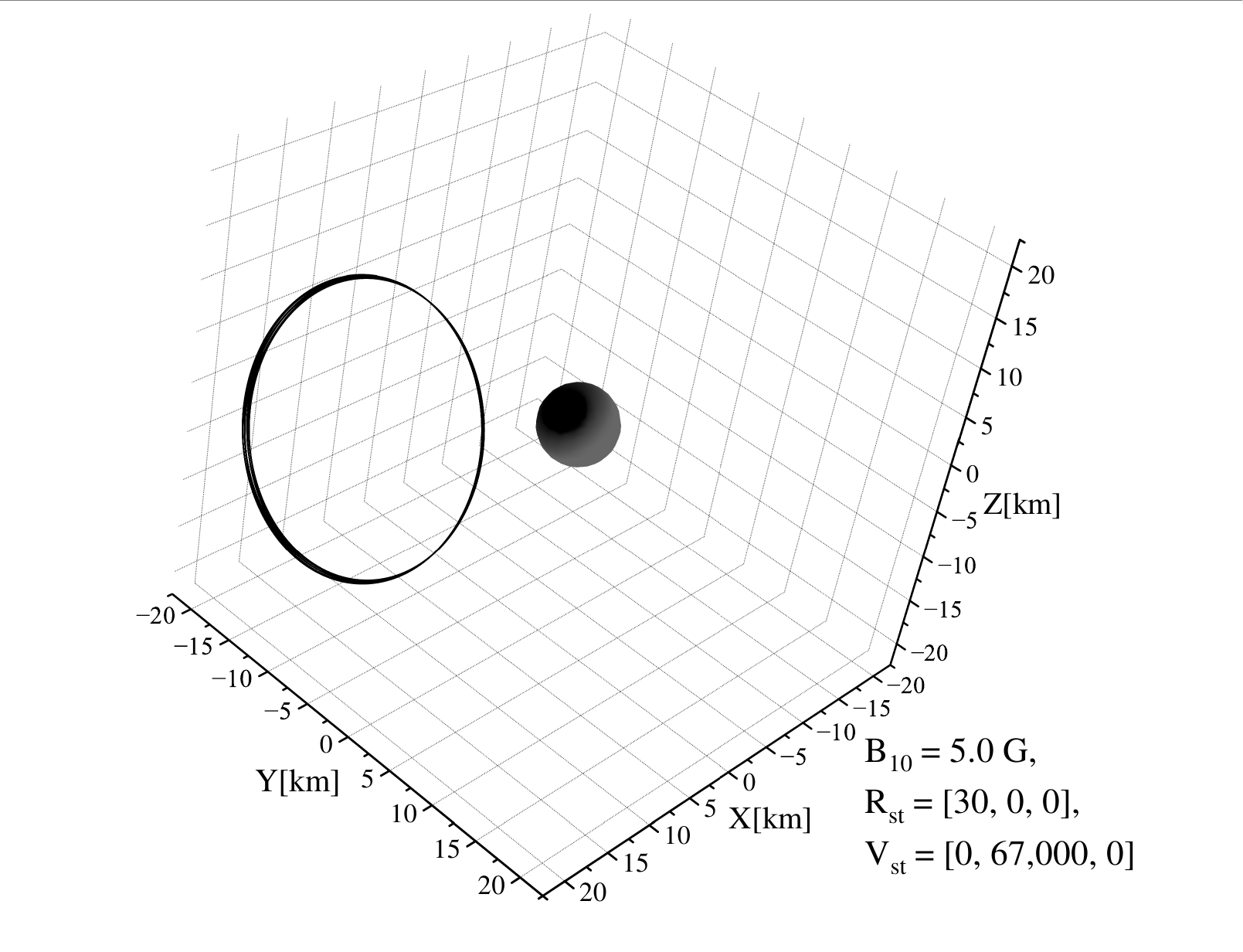}
  \caption{{Trajectories} (black line) 
 of a proton near a point-like gravitating object (black sphere) in the fictional space with the mass of the Sun and a monopole magnetic field for different starting parameter values. $B_{10}$ is the magnetic field strength at $R = 10$km, and $R_\text{st}$ and $V_\text{st}$ are the starting position and speed of a particle in km and km/s, respectively.}
  \label{fig01}
\end{figure}

Figure \ref{fig01} shows some of these trajectories. We can see that even a relatively weak ($B_g \sim 1$G) monopole magnetic field prevents, as we conjectured earlier, the emergence of classical Keplerian orbits around the object. Instead, the proton begins to form spiral trajectories around radial magnetic field lines. The form of this spiral and the direction of the proton movement strongly depend on the parameters. In particular, at $\sim$67,
000 km/s, the proton’s trajectory becomes closed (see the bottom picture in Figure \ref{fig01}). This is a kind of ``first escape velocity'' for this particular situation. However, it should be noted that if we take into account the loss of energy due to cyclotron radiation, such circular orbits would not be able to exist for long and would turn into spiral ones. Accordingly, at speeds lower than $\sim$67,000~km/s, the proton moves toward the central object, and at speeds greater than $\sim$67,000~km/s, it moves away from the central object. As the magnetic field strength increases, the radius of the helix quickly decreases, essentially leading to the almost radial motion of the proton.

\section{Conclusions}

The study of accretion into wormholes in the presence of a monopole magnetic field could have important consequences for fundamental physics and cosmology. Observations of accretion flow near wormholes could provide insights into the nature of dark matter, the origin of high-energy cosmic rays, and the nature and properties of compact objects in active galactic nuclei. Also, the study of wormholes and magnetic fields offers a unique opportunity to test the limits of general relativity and explore exotic spacetime geometries beyond the predictions of classical physics, offering a way to explore the interaction between gravity, magnetism, and spacetime geometry. Theoretical models, numerical simulations, and astronomical observations of hypothetical accreting wormholes can provide valuable information about the physical parameters, dynamics, and observable signatures of these interesting astrophysical~objects.

\textls[-25]{In the current paper, we have shown that a wormhole with a monopole magnetic field can generate non-polarized cyclotron radiation, which is unusual for known astrophysical~objects.}

Possible candidates for such objects include both supermassive relativistic objects in the centers of galaxies and primordial wormholes of medium and small ``mass'' formed in the early Universe. In particular, the latter may appear as star-like objects with an unusual non-polarized non-thermal spectrum consisting of cyclotron or synchrotron emissions.

Calculations based on a toy model suggest that even a relatively weak ($\sim$1 G) monopole magnetic field prevents the emergence of classical Keplerian orbits around the object, which justifies our conjecture about the mainly radial character of accretion at a fairly close distance from the wormhole. However, the formation of an accretion disk is still quite possible at some distance from the wormhole.

Also, if we consider the case of the accretion disk, a monopole magnetic field could complicate its formation near the wormhole and lead to the emergence of asymmetrical and one-sided relativistic jets.

Future space missions and ground-based facilities will play important roles in advancing our understanding of magnetized accretion into relativistic objects and addressing open questions in modern astrophysics. Space observatories, such as the James Webb Space Telescope, the Nancy Grace Roman Space Telescope, and the European Space Agency's Athena mission, will significantly improve sensitivity and wavelength coverage for studying magnetized accretion flows. Ground-based facilities such as the Atacama Large Millimeter/Submillimeter Array, the Square Kilometer Array, and the upcoming Giant Magellan Telescope and Extremely Large Telescope will allow astrophysicists to probe magnetized accretion with high angular resolution and sensitivity.

\vspace{6pt}

\authorcontributions{Conceptualization, M.P. and S.K.; methodology, M.P. and S.K.; validation, M.P. and S.K.; formal analysis, M.P., S.K., S.B., and T.N.; investigation, S.B. and T.N.; resources, S.B. and T.N.; data curation, S.B. and T.N.; writing--original draft preparation, M.P., S.K., and S.B.; writing--review and editing, M.P. and S.K.; visualization, M.P.; supervision, M.P. and S.K.; project administration, M.P. All authors have read and agreed to the published version of the manuscript.}

\funding{{This} 
 research was supported by the state order of the Central Astronomical Observatory at Pulkovo {as part of the planned research topic for ``MAGION''---Physics and Evolution of Stars and Active Galactic~Nuclei.}
 }

\dataavailability{{Data} are contained within the article. 
}


\conflictsofinterest{The authors declare no conflicts of interest.}

\begin{adjustwidth}{-\extralength}{0cm}

\reftitle{References}

\PublishersNote{}
\end{adjustwidth}
\end{document}